\documentclass[journal,transmag]{IEEEtran}
\usepackage{cite}
\usepackage{graphicx}
\usepackage{times}
\usepackage{latexsym}

\usepackage[mathscr]{eucal}
\usepackage{array}
\usepackage{fancyhdr}
\usepackage{amsmath,amssymb}
\usepackage{bm} 
\usepackage{citesort}
\usepackage{multirow}

\ifCLASSINFOpdf
\else
\fi

\usepackage{xcolor}

\makeatother

\usepackage{cite}
\usepackage{amsmath,amssymb,amsfonts}
\usepackage{algorithmic}
\usepackage{graphicx}
\usepackage{textcomp}
\usepackage{diagbox}
\usepackage{xcolor}
\usepackage{booktabs}
\usepackage{xspace}
\usepackage{float}
\usepackage[nameinlink]{cleveref}
\usepackage[group-separator={,}, group-four-digits, per-mode=symbol, binary-units]{siunitx}

  \usepackage[caption=false,font=footnotesize]{subfig}

\ifCLASSINFOpdf
\else
\fi

\begin{document}

\title{Information-Theoretic Secure Key Sharing for Wide-Area Mobile Applications}
\author{Guyue Li,~\IEEEmembership{Member,~IEEE}, Hongyi Luo, Jiabao Yu, Aiqun Hu,~\IEEEmembership{Senior Member,~IEEE} and Jiangzhou Wang,~\IEEEmembership{Fellow,~IEEE}
    \thanks{Guyue Li and Hongyi Luo are with the School of Cyber Science and Engineering, Southeast University, Nanjing 210096, China (e-mail: guyuelee@seu.edu.cn; hongyiluo@seu.edu.cn).}
    \thanks{Jiabao Yu is with the Purple Mountain Laboratories, Nanjing 210096, China (e-mail:yujiabao@pmlabs.com.cn).}
    \thanks{Aiqun Hu is with National Mobile Communications Research Laboratory, Southeast University, Nanjing 210096, China (e-mail: aqhu@seu.edu.cn).}
    \thanks{Jiangzhou Wang is with the School of Engineering, University of Kent, Canterbury CT2 7NT, U.K. Email: (e-mail: j.z.wang@kent.ac.uk).}
    \thanks{Guyue Li and Aiqun Hu are also with Purple Mountain Laboratories, Nanjing 210096, China.}
 }
\maketitle
\begin{abstract}
With the rapid growth of handheld devices in the internet of things (IoT) networks, mobile applications have become ubiquitous in everyday life. As technology is developed, so do also the risks and threats associated with it, especially in the forthcoming quantum era. 
Existing IoT networks, however, lack a quantum-resistant secret key sharing scheme to meet confidential message transmission demands in wide-area mobile applications. 
To address this issue, this article proposes a new scheme, 
channel reciprocity (CR) based quantum key distribution (QKD) 
CR-QKD, which accomplishes the goal of secret key sharing by combining emerging techniques of QKD and CR-based key generation (CRKG).
Exploiting laws of quantum physics and properties of wireless channels, the proposed scheme is able to ensure the secrecy of the key, even against computationally unbounded adversaries.
The basic mechanism is elaborated for a single-user case and it is extended into a multi-user case by redesigning a multi-user edge forwarding strategy. In addition, to make CR-QKD more practical, some enhancement strategies are studied to reduce the time delay and to improve the secret key generation rate in a secure manner. A prototype of CR-QKD is demonstrated in a metropolitan area network, where secret keys are shared between two remote IoT devices that are roughly 
fifteen kilometers apart from each other. The experimental results have verified that CR-QKD allows a secret key rate of 424 bits per second with a retransmission rate of 2.1\%.
\end{abstract}

\begin{IEEEkeywords}
Secret key generation, physical layer security, quantum key distribution, wide-area mobile applications, Internet of Things
\end{IEEEkeywords}

%

%

\section{Introduction} \label{sec:introduction}
Recent years have witnessed a remarkable growth in the number and variety of mobile devices and applications in the Internet of Things (IoT) networks. The flourish of IoT, however, has resulted in the generation of a substantial amount of private messages exchanged over public channels, which has grabbed one's attention. Unfortunately, IoT devices are susceptible to various threats and security challenges, which pose hazards for the advancement of IoT in sensitive fields, such as smart homes, unmanned vehicles, e-health, and military networks~\cite{iot2021}. 
In order to avoid being revealed to a third party, a message is usually encrypted using a secret key shared among the communicating devices. Thus, a key prerequisite of achieving IoT network security is secret key sharing that avoids eavesdropper interception~\cite{end2end2016}. 
   
In a classic cryptographic scheme, two legitimate parties, namely Alice and Bob use the public-key cryptosystem (PKC) for key distribution. It is extremely difficult for a third party, namely Eve, to derive the private key or message computationally, due to the intractability of certain mathematical problems used in encryption algorithms.  
However, the emerging quantum computing technology has the potential to make some previously-intractable problems tractable~\cite{qc2013}. 
Thus, the security of computational security-based key distribution will be rendered insecure by substantial progress in quantum computing in the coming years, which necessitates the study of alternative solutions that do not rely on computational security. 


\begin{table*}[htbp] \label{table1}
\centering
\caption{A Summary and Comparation of Typical Secret Key Distribution Methods}
    \begin{tabular}{|c|c|c|c|c|c|} 
        \hline 
        \diagbox{Method}{Metric}& Security Level& Mobility Support & Distribution Distance & User Cost \\ \hline
        PKC& Computational secure& Middle&\textbf{Long}& Middle    \\ \hline
        QKD&\textbf {Information theoretically secure} & Weak&\textbf{Long}&High\\ \hline
        CRKG& \textbf{Information theoretically secure} &\textbf{Strong}& Short&\textbf{Low}\\ \hline
        \textbf{CR-QKD}& {\textbf{Information theoretically secure}} &{\textbf{Strong}}&{\textbf{Long}}&{\textbf{Low}}  \\ \hline      
    \end{tabular}
 \end{table*}
In this context, much attention has been paid to emerging techniques, such as quantum key distribution (QKD)~\cite{Q6G2022} and channel reciprocity-based key generation (CRKG)~\cite{Li2019Physical}, which can provide secret key sharing service with information-theoretic security, also known as unconditional security or physical security. 
\begin{itemize}
\item QKD is a well-known quantum-resistant mechanism, which distributes secret keys to distant parties by transmitting single photon through a quantum channel~\cite{2013Nature}. Employing the laws of quantum physics, QKD can detect eavesdroppers during the key generation process, in which unauthorized observation of quantum communication induces a discernible increase of errors. This sensitivity to eavesdropping makes QKD possible to ensure the secrecy of the key, even against computationally unbounded adversaries. 
\item CRKG is built on the basis of channel reciprocity, which means that the channel responses of the forward and backward communication links are very similar in a time division duplex (TDD) system. In addition, the dynamic and complex wireless communication environment makes the channel responses change over time and hard to predict. Therefore, legitimate users can share a pair of common randomness from their radio channel measurements. 
Since CRKG does not require assistance from a third party nor expensive infrastructure, it has recently emerged as a new paradigm that provides a lightweight and information-theoretic secure key sharing solution for decentralized or device-to-device sensor applications~\cite{maurer1993secret}. 
\end{itemize}

Table~\ref{table1} summarizes these typical secret key distribution methods, and identify their characteristics from perspectives of security level, mobility support, distribution distance and user cost. 
We find that although the separate construction of QKD and CRKG can be supported in the physical layer, there is no investigation of a secret key sharing scheme for the security demands from remote mobile devices. Although point-to-point connections are suitable to form a backbone quantum core network to bridge long distances, they are less suitable to provide the last-mile service needed to give a multitude of users access to this QKD infrastructure~\cite{2013Nature}. Similarly, despite many research efforts in the field of CRKG, its widespread application is unfortunately hindered by the short distance between transceivers. 
With a rapid growth of handheld devices, wide-area mobile applications, such as remote environmental and elderly monitoring, have become an inseparable part of IoT networks. 
A new architecture needs to be developed where end-users between two access networks are connected to a metro network, thus realizing unconditionally secure key sharing in a more cost-effective and flexible manner. 
 
In this article, we introduce and experimentally demonstrate the concept of a ‘channel reciprocity-aided quantum key distribution (CR-QKD)’ based on simple and cost-effective telecommunication technologies. 
This scheme can expand the scope of QKD to IoT networks and therefore vastly broaden users' appeal. 
The contributions of this article are three-fold:
\begin{itemize}
    \item We introduce a novel secret key sharing architecture, referred to as CR-QKD, which bridges a backbone quantum core network and IoT users by exploiting the technique of CRKG to provide the last-mile service. CR-QKD is information-theoretically secure and it does not require IoT users to be equipped with expensive quantum infrastructures for exchanging secret keys, thereby significantly reducing the hardware requirements.
     \item We propose a multi-user mechanism to realize the concept of CR-QKD with an elaborate design of key alignment. We also identify challenges that arise due to the hybrid architecture of CR-QKD from the perspective of feasibility and security, respectively. Countermeasures have been studied to reduce the time delay and to improve the secret key generation rate in a secure manner. 
     \item We implement a prototype CR-QKD system in a metropolitan area network, in which secret keys are shared between two remote IoT devices that are roughly fifteen kilometers apart from each other. The experimental results have verified that CR-QKD can provide a secret key rate of 424 bits per second with a retransmission rate of 2.1\%.
\end{itemize}

 \begin{figure*}[htbp]
    \centering
    \includegraphics[height=4.8cm,width=12cm]{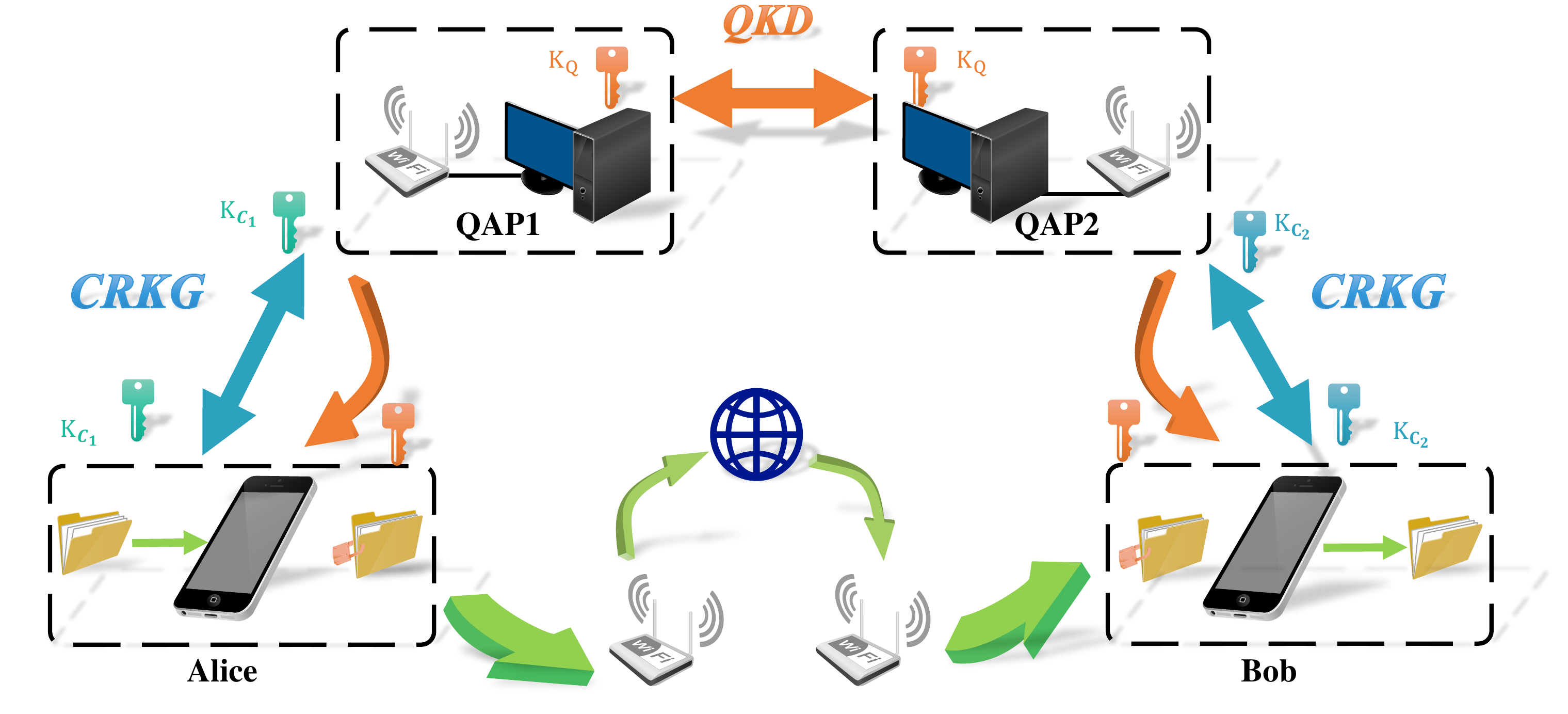}   
     \caption{An illustration of combining QKD and CRKG to realize secure communication between two remote users.}
    \label{mode}
  \end{figure*}

\section{An Overview of the CR-QKD architecture}
In this section, we first introduce QKD and CRKG, and then discuss their combination modes to realize secure communication in wide-area mobile applications. 
\subsection{QKD}
QKD protocols exploit a quantum communication channel and an authenticated classical channel to ensure the exchange of a cryptographic key between two remote parties with proven security. Since its inception in~\cite{BB84}, QKD protocol design and analyses have flourished as a field yielding numerous protocols, security analyses, and practical implementation methodologies. 
Although QKD research has made remarkable progress, these developments have been largely focused on securing large-scale infrastructures using long distance fiber transmission and free space transmission between fixed terminals. 
Some efforts have been made toward handheld free-space QKD by exploiting a beam-steering module, which compensates for hand movement of the QKD module at the transmitter~\cite{Chun17,Elmabrok2018Wireless}. However, these schemes have limited transmission range and their QKD receiver is currently difficult to be miniaturized. 
In other words, they can not provide a bi-directional transmission and are thus not applicable to the case of distributing a quantum key from a core network to an end-user. In this article, QKD is exploited to form a backbone quantum core network to bridge long distances.
\subsection{CRKG}
CRKG exploits wireless channels between transceivers as random sources for key generation, and these keys can be replenished dynamically as wireless channels vary over time. Eavesdroppers in such situations experience physical channels independent of those of the legitimate users as long as they are a few wavelengths away from these legitimate parties, which is generally the case in wireless networks. 
So far, the CRKG field has yielded fruitful results from aspects of theoretical exploration, modeling, protocol design, and prototype implementation in various IoT platforms~\cite{ZHANG2020Frontier}.
However, these developments have been largely focused on wireless communication technologies for short-range applications, such as ZigBee, ultra-wideband, Bluetooth and WiFi. When the distance is in the order of a few kilometers, the signal-to-noise ratio is small and the time delay between uplink and downlink packets becomes large. Therefore, CRKG at a long distance is challenging to meet the requirement of high correlation between channel parameter measurements for secret key generation~\cite{ZHANG2020Frontier}.
Due to these reasons, CRKG is more suitable for secret key sharing between wireless transceivers that are within one kilometer apart and thus exploited in this article to complete the last-mile secret key distribution task from quantum access points (QAP) to IoT users.

 
  

\subsection{The Combination Mode of QKD and CRKG}
Neither QKD nor PKG is applicable to long-range IoT networks, therefore, a critical problem is how to combine their advantages to apply to the new scenario. Fig.~\ref{mode} describes the system model and illustrates one possible combination mode. 
Alice and Bob are two distant wireless users, who do not have direct links with each other. QAP1 and QAP2 are two quantum nodes that are connected through long-distance optical fibers, or ground-to-satellite free-space links. QAP1 and QAP2 have a wireless link to Alice and Bob, respectively.  
  
To complete the secret key distribution between Alice and Bob, three keys are first shared between Alice and QAP1 (link 1), QAP1 and QAP2 (link 2), and QAP2 and Bob (link 3).
Channel keys are generated from wireless links 1 and 3 by using the technique of CRKG, while quantum key is distributed from QAP1 to QAP2, or in the reverse direction, with mature QKD techniques. 
Next, the quantum key is securely delivered to Alice and Bob by encrypting it with channel keys. In other words, Alice and Bob share a unified key, which is then used to encrypt and to decrypt the message in the data transmissions.
Therefore, this mode is also abbreviated as unified-key mode. 
Notably, Alice and Bob are free to choose wireless and Internet routes for message transmission. This consideration is due to the following reasons. First, due to the limited rate of the quantum link, its message transmission rate is relatively small. 
Second, as Alice and Bob are mobile devices, they are more likely to use communication routes that are different from those in the key distribution process. Finally, the unified-key requires less time delay for message transmission as it only needs one time of message encryption and decryption. 
The essential process to obtain unified quantum keys is referred to as CR-QKD, which is elaborated in the following section.
  \begin{figure*}[htbp]
    \centering
    \includegraphics[height=6cm,width=17.2cm]{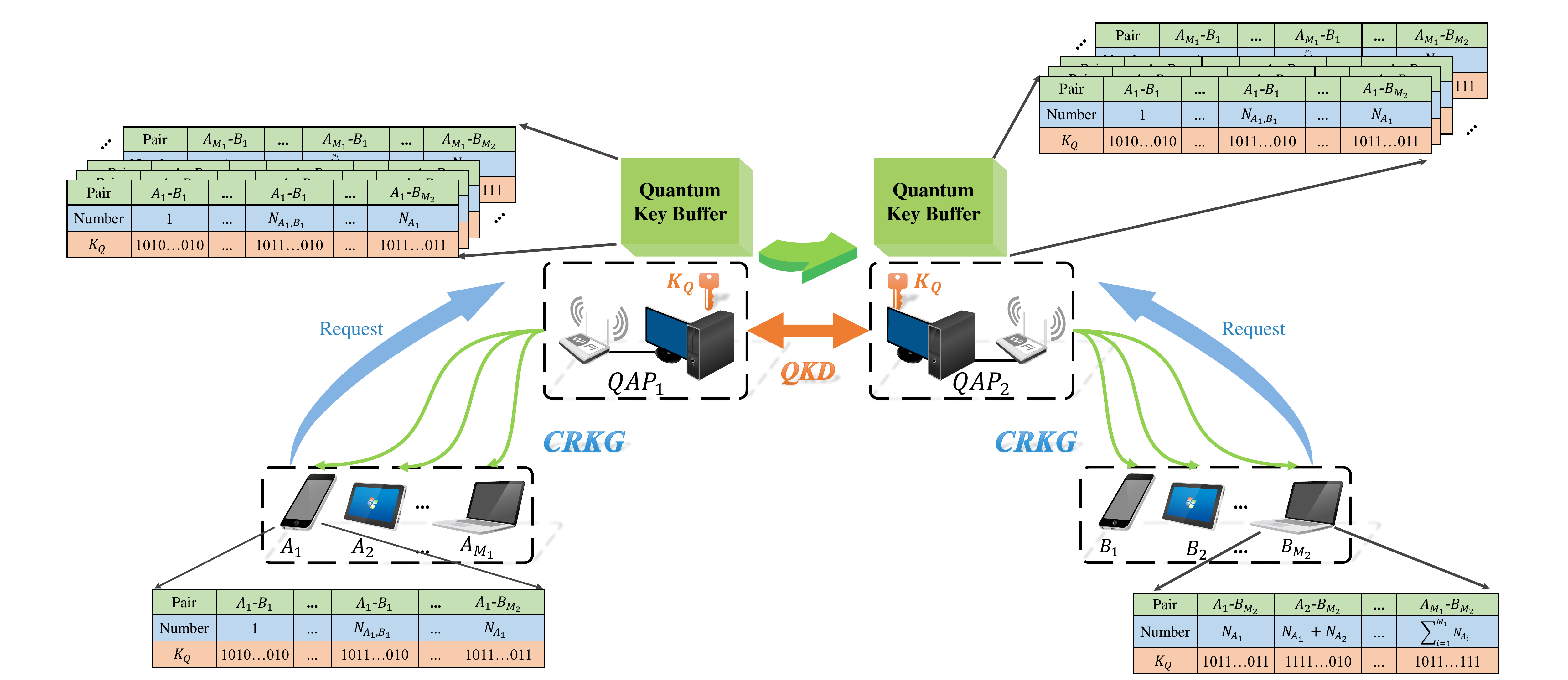}
    \caption{An illustration of CR-QKD in a multi-user scenario: edges distribute quantum keys to users according to their needs, where $N_{A_i,B_j}$ represents the number of key groups required by $A_i-B_j$ user pair and ${N_{A_i}}$ represents the total number of key groups required by user $A_i$.}
    \label{mutiuser}
  \end{figure*}  
\section{Conceptual design of CR-QKD}
In this section, we will first introduce the basic mechanism of CR-QKD and then study the key aligment, efficiency and security issues that exist in CR-QKD.
\subsection{Mechanism description}\label{sec:IIIA}
As shown in Fig.~\ref{mode}, Alice and Bob intend to share quantum key with the help of QAP1 and QAP2, against an adversarial eavesdropper, Eve, tapping on the quantum channel and listening to all the exchanges on the classical channels. 
Similar to most existing QKD and CRKG protocols, the classical communication channels are assumed to be authenticated, in which the identities of the communicating parties have been verified and the integrity of the transmitted messages is promised.

The CR-QKD protocol comprises three main phases, i.e, QKD~\footnote{Our study is not bound to specific QKD protocols, and we choose the BB84 protocol as a representative to introduce the CR-QKD mechanism.}, CRKG and edge forwarding, which will be elaborated below.
\begin{itemize}
\item {\textbf {QKD phase:}} First, QAP1 prepares and sends to QAP2 a set of random qubits via a single-photon signal over a quantum channel. These qubits are selected from a set of four states with two bases. 
For every incoming state from QAP1, QAP2 randomly chooses one of the two bases to measure and record the results. 
Once quantum communication has finished, QAP2 starts base reconciliation by announcing the position of the detected bits and the basis used to QAP1 over a classic channel. Then, QAP1 and QAP2 retain the bits with a coincident basis and discard the rest. 
After that, QAP1 publishes a subset of these bits to QAP2 for eavesdropping detection. 
If the error rate between what QAP2 detects and what QAP1 has sent is high, the eavesdropping is detected and these shared bits will be invalid. 
Otherwise, QAP1 and QAP2 perform information reconciliation and privacy amplification over the rest of the bits that have not been made public. At last, QAP1 and QAP2 check whether they obtain the same result via key verification. If so, they retain the pair of bits as quantum key $K_Q$, otherwise, they discard both of them. 
\item {\textbf {CRKG phase:}} 
A CRKG protocol typically contains four stages,
i.e., channel probing, quantization, information reconciliation, and privacy amplification.
Alice and QAP1 first carry out channel probing, which involves bidirectional measurements within a channel coherence time. They then convert the analog measurements into digital binaries.
There will probably be a mismatch between these binaries, hence information reconciliation has to be adopted to correct the mismatch. To avoid information leakage, privacy amplification is employed to distill the reconciliated binaries. Finally, after key verification, Alice and QAP1 retain the pair of bits as channel key $K_{C1}$ and Bob and QAP2 retain the pair of bits as channel key $K_{C2}$.
\item {\textbf {Edge forwarding phase:}}
In the last phase, previous quantum keys shared between QAP1 and QAP2 are forwarded to Alice and Bob, completing the ultimate task of secret key sharing.  
Security is the primary concern here, as eavesdroppers should not learn any information about the quantum key through this forwarding process. With the help of channel keys, it is possible for edges to encrypt quantum keys with them using the One-Time-Pad (OTP) encryption algorithm and then forward the ciphertext to users. So far, the secret key sharing task is completed.
\end{itemize}
Although CR-QKD provides a potential solution, it still faces some challenges to be implemented in practice. We divide these challenges into three categories and discuss 
along countermeasures below. 
\subsection{Key alignment}
OTP is a well-known example of encryption scheme that provides ``perfect secrecy", however, one challenge here is that the channel key used for OTP must be at least as long as the quantum key to be encrypted. As channel keys, $K_{C1}$ and  $K_{C2}$, are generated from different wireless channels, their key generation rates are likely to be different from each other, and that of the quantum key $K_{Q}$. 
As a result, the quantum keys distributed to Alice and Bob may be disordered.  

We address the key alignment issue by segmenting quantum key and channel key into groups and numbering them before edge forwarding.  
Each group has a fixed bit number of $L_G$. Those quantum key and channel key bits belonging to the same group are encrypted through a binary XOR operation. 
Then, the ciphertext is forwarded, together with the group number. 
Alice and Bob eventually obtain the quantum key by decrypting the ciphertext using their corresponding channel keys. Here, the trade-off between overhead and real-time must be taken into account in the selection of the group size.  If the group size is small, the group number will occupy a field length comparable to that of the ciphertext, and the communication overhead will become significant. Otherwise, if the group size is large, the communication overhead is reduced but it will take a long time to accumulate sufficient keys for forwarding.

Next, we extend the key alignment issue into a multi-user scenario, where $A = \{A_1, A_2, \cdots, A_{M_1}\}$ and $B = \{B_1, B_2, \cdots, B_{M_2}\}$ are two sets of IoT users at the service range of QAP1 and QAP2, respectively. Users in $A$ desire to share secret keys with users in $B$. When CR-QKD is applied to this case, a new problem arises, i.e., how to distribute quantum keys from the edge to multiple users, who have different requirements and channel conditions. 
In this article, we introduce a multi-user edge forwarding strategy, which distributes quantum keys to each user according to its needs. Fig.~\ref{mutiuser} illustrates one round of the quantum key distribution process using this strategy. 

To start with, users in $A$ broadcast the name of their target users for key sharing and the number of required key groups. 
After receiving these requests, QAP1 shares the information with QAP2. Then, QAP2 broadcasts it over the air and the relevant users in $B$ record them locally.
Next, quantum key sequences are shared between QAP1 and QAP2 through the above QKD phase. These quantum key sequences are segmented into groups and numbered, each having $L_G$ bits. QAP1 allocates quantum key groups for each user pair according to their requests. 
The mapping relationship of user pairs and the key group number is transmitted to QAP2. This allocation information is saved in a quantum key buffer. In this way, the quantum keys are synchronized at QAP1 and QAP2. 
Next, they yield channel keys with these demanding users, respectively. 
For each user, the CRKG process is performed multiple times until it has accumulated sufficient number of key groups.
Finally, QAP1 and QAP2 use these CRKG keys to encrypt the corresponding quantum keys and broadcast the ciphertext together with the user pairs and group number to end-users. Each end-user obtains quantum keys by decrypting the related ciphertext with its own CRKG keys. 
Finally, these quantum keys are divided into each user pair for message encryption and this round of quantum key distribution has come to an end. 


\subsection{Efficiency improvement}
The basic CR-QKD mechanism is time-consuming as it interacts heavily to obtain identical keys in both QKD and CRKG phases. This situation becomes more severe in a multi-user case. For each round of multi-user key distribution, in a time division multiple access (TDMA) system, the time delay is the sum of the time spent on yielding quantum keys and channel keys plus the time used for key forwarding. The time spent on quantum keys is calculated by dividing the number of quantum key bits by the quantum key generation rate.
The time spent on channel keys is equal to the larger one of QAP1 and QAP2.
For each QAP, its time delay is the sum of that used for yielding channel keys between it and all users.
One approach to reducing the time delay is to make QKD and CRKG processes work in parallel. However, its reduction ratio is less than 50\% due to the positive forwarding time and the maximum operation. 

Another solution to further reduce the time delay is to improve the secret key generation rate.
In practice, key generation rates are largely subject to the long time delay caused by information reconciliation, which exchanges parity information or syndromes over classic channels to detect and correct errors in the preliminary key material. According to OTP with un-identical keys~\cite{OTP}, we propose a simplified CR-QKD mechanism that abolishes the sophisticated information reconciliation step in the CRKG phase and forwards quantum keys using non-reconciled channel keys. 
The challenge is to decrypt the quantum keys correctly when the non-reconciled channel keys of two parties are different but highly correlated. We deem the XOR encryption and decryption modules along with the physical channel as an equivalent cascade channel. Then, the tiny differences between keys can be seen as part of the transmission error, and thus can be corrected by the off-the-shelf channel coding with a stronger correction capability. 
\begin{figure}[htbp]
  \centering
  \includegraphics[height=7cm,width=9.31cm]{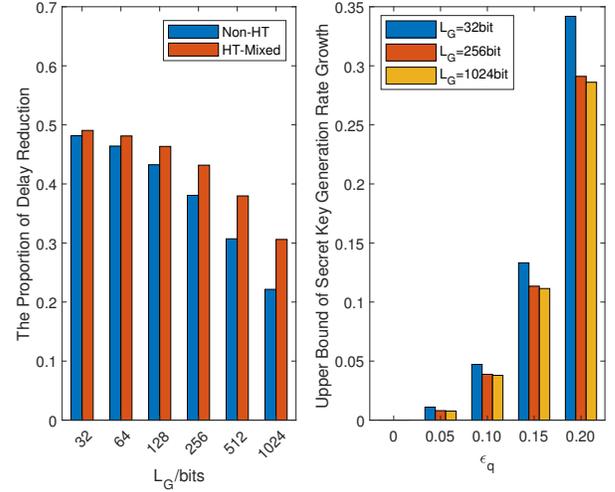}
  \caption{Performance improvements of time delay and secret key generation rate in a typical WiFi scenario: the transmission distance is set as 150 meters and the bandwidth is set as 20 MHz. The fixed overhead of a WiFi frame under the Non-HT (Non-High Throughput) and HT-Mixed mode is $20$ us and $40$ us, respectively.
  }
  \label{Communication Consuming Proportion}
\end{figure} 
\begin{figure*}[htbp]
  \centering
  \includegraphics[height=5.5cm,width=16cm]{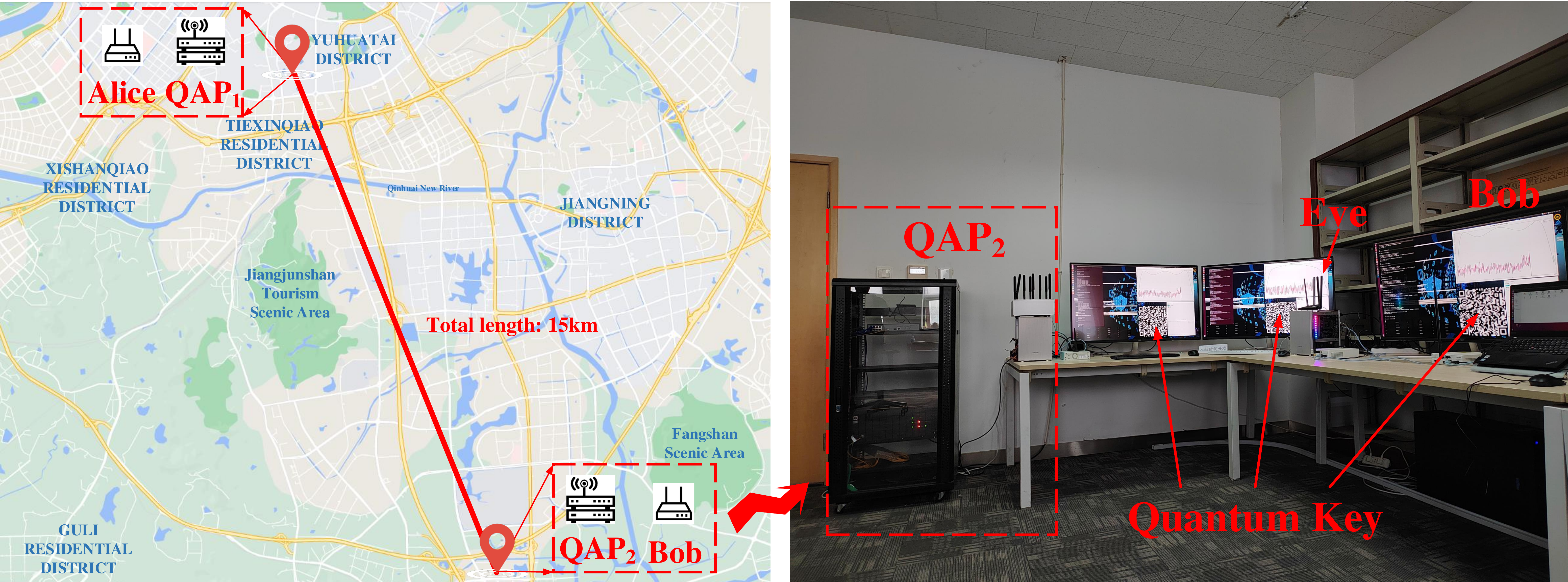}
  \caption{An illustration of the CR-QKD prototype platform in a metropolitan area network at Nanjing, which is the capital of Jiangsu Province, East-central China. One is located in Yuhuatai District and the other is at Chinese Network Valley in Jiangning District.}
  \label{map}
\end{figure*} 
Fig.~3 plots performance improvement ratios of the simplified CR-QKD mechanism compared with the paralleled CR-QKD mechanism in terms of time delay and upper bound of secret key generation rate in a typical WiFi scenario. As shown in the left panel, the proportion of delay reduction decreases with the rise of $L_G$, still achieving a reduction ratio above 20\% at $L_G \le 1024$. The reduction of HT-Mixed mode is more remarkable than Non-HT mode, as the former has a larger time overhead than the latter. The right panel shows that the growth of the upper bound of the secret key generation rate is more remarkable when the bit disagreement ratio between quantized channel measurements gets larger, while it has a slight fall with the rise of  $L_G$. When $L_G =1024$ and $\epsilon_q =0.1$, the proportion of delay reduction and upper bound of secret key rate growth are roughly 20\% and 10\%, respectively. These simulation results verify the effectiveness of the proposed simplified CR-QKD mechanism.
  
\subsection{Security enhancement}
Another challenge of CRKG lies in the increased security risks caused by its hybrid architecture, as security is only as strong as its weakest link. We assume that the terminal security of QAP1 and QAP2 is guaranteed by techniques such as trusted computing. Operations that are relevant to secret keys are run in a trusted execution environment, thereby attackers can read neither quantum keys nor channel keys from the hybrid interface on QAP1 and QAP2.  Since the edge forwarding phase employs the OTP encryption scheme, its security depends on the key used for OTP.
The security of existing CRKG approaches, however, heavily relies on the channel variation and thus suffers from vulnerabilities in slowly varying environments~\cite{2020Physical}. When users have low mobility, e.g. in a wireless sensor network, there exist inevitable and unknown temporal correlations between adjacent channel samples, resulting in a large proportion of repeated bit segments in the quantized bit sequences. 
Several solutions can be used to facilitate the practical usage of CRKG in slowly varying environments.
One solution is to introduce helper devices, e.g., relays and reconfigurable intelligent surface (RIS) to boost the key generation rate and randomness~\cite{WCM}. However, this solution encounters some practical problems, such as the unavailability of trust relays and
additional hardware overheads of RIS devices. 
Another idea is to scramble these bits segments through some permutation or interleaving techniques. However, the security of the key may be compromised when the permutation information is public. \cite{INFOCOM} has proposed a new physical-layer secret key generation approach with channel obfuscation, which improved the dynamic property of channel parameters based on random filtering and random antenna scheduling, which have mutual remedying parameters in hiding the obfuscation information.
   
\section{Case study: An Implementation of CR-QKD}
To realize the concept of CR-QKD, we implement a single-user confidential transmission prototype system in a metropolitan area network. 
\subsection{Experimental Setup}
As shown in the left panel of Fig.~\ref{map}, QAP1 and QAP2 are two quantum access points at a distance of fifteen kilometers. Alice and Bob are two remote IoT users in the wireless service ranges of QAP1 and QAP2, respectively. Without loss of generality, we zoom in on the wireless access network at Chinese Network Valley, as depicted in the right panel of Fig.~\ref{map}. 
Here, QAP2 is composed of a QKD terminal under the series of QKDM-POL40-S for yielding quantum keys~\footnote{The quantum keys meet strict key randomness, as they conform to the specification of the GM/T 0005-2012.}, a USRP N210 SDR device embedded with the CBX daughterboards for providing a wireless connection service, and a computer under the trusted execution environment for yielding wireless channel keys and distributing quantum keys. Both the QKD terminal and USRP N210 are connected to the computer via the ethernet cable in QAP2. 
The end-user, Bob, and a passive eavesdropper, Eve, are realized through two USRP N210 SDR devices, respectively. We design a TDD frame for channel sounding, which consists of a sinusoidal sequence for synchronization and an M-sequence for channel estimation. The signal operates at 2.605GHz and 20MHz bandwidth to avoid collisions with ubiquitous 2.4GHz signals such as WiFi. Once Bob receives the channel sounding signal from AP2, it will immediately switch to TX mode and send the same channel sounding signal. By using the same channel sounding signal for channel estimation, the amplitude part of the CSI is further preprocessed and quantized to generate the wireless keys.
\subsection{Performance Results}
Considering the comparable experimental scenarios and results of QAP1 and QAP2, we only take QAP2 as an example for performance analysis. 

Table~\ref{table2} summarizes the secret key sharing results from QAP2 to Bob and Eve in three typical indoor scenarios, namely office, hall and corridor. First, the measured key generation rates (KGRs) of the channel keys between QAP2 and Bob in above scenarios are 315.4, 424.7 and 383.7 bits per second (bps), respectively. They are sufficient for traditional symmetric encryption algorithms (such as AES) to update 256-bit keys every second for secure communications. In the random test, we examined a bit sequence of length 3.4 million bits that was obtained at the output of the quantization stage without further processing. 
The generated channel keys passed 14 NIST statistical tests, indicating their randomness. 
However, while the simplified CR-QKD mechanism leads to high KGRs and high randomness, removing the complicated information reconciliation step also results in relatively high key disagreement rates (KDRs) of 8.1\%, 4.7\% and 5.8\% between QAP2 and Bob, respectively. 
The number of personnel, the frequency of movement, and the switching time of USRP 
affect the reciprocity of uplink and downlink channels, eventually leading to KDR differences in the noisy office, occasionally infested corridor, and empty hall.
Meanwhile, along with forwarding quantum keys using non-reconciled channel keys based on channel error correction coding, the need arises to retransmit quantum keys when unsuccessfully decoded. 
The corresponding retransmission rates (RRs) using Polar codes from QAP2 to Bob are 11.6\%, 2.1\%, and 6.7\%, respectively, which are proportional to the KDRs. 

To demonstrate the security of our proposed scheme, we also evaluate the quantum key cracking performance of the near-end eavesdropper Eve in terms of KDR and cracking rate (CR).
The KDRs between QAP2 and Eve under these three scenarios are all around 50\%, where the line of sight in the straight corridor contributes to a relatively lower KDR but is still above 45\%. What's more, the experimental results show that the CRs of Eve in the three scenarios are all zero, which means that none of the quantum keys have been cracked.

\begin{table}[htbp]
\caption{The Quantum Key Wireless Distribution Performance in Three Indoor Scenarios} \label{table2}
\centering{}%
\begin{tabular}{|c|c|c|c|c|c|c|}
\hline 
Scenario & \multicolumn{4}{c|}{QAP2 - Bob} & \multicolumn{2}{c|}{QAP2 - Eve}\tabularnewline
\hline 
Metrics & KGR/bps & NIST & KDR  & RR & KDR & CR\tabularnewline
\hline 
Office & 315.4 & 14 & 8.1\%  & 11.6\% & 48.1\% & 0\%\tabularnewline
\hline 
Hall & 424.7 & 14 & 4.7\%  & 2.1\% & 49.2\% & 0\%\tabularnewline
\hline 
Corridor & 383.7 & 14 & 5.8\% & 6.7\% & 45.3\% & 0\%\tabularnewline
\hline 
\end{tabular}
\end{table}

\section{Conclusion and Future Directions}
Integrating QKD into IoT networks is beneficial for QKD’s practical deployment and end-user’s security enhancement. This article proposed a framework of CR-QKD over IoT networks. QKD and CRKG assembly were adopted for secret key sharing over backbone core networks and the last-mile wireless access networks in CR-QKD, respectively. 
The demonstration of CR-QKD prototype represented a major step towards real-world information theoretically security for wide-area mobile applications, such as confidential VoLTE and confidential VoWiFi.

Some open issues in future work are given below.
\begin{itemize}
    \item \textbf{Device Authentication}: Considering the hybrid architecture of CR-QKD, it is more vulnerable to spoofing attacks from either user's side or QAP's side. However, neither QKD nor CRKG provides a means to authenticate the transmission source. Therefore, source authentication in CR-QKD should be further studied by using asymmetric cryptography techniques or emerging physical-layer techniques, such as radio frequency fingerprinting identification and physical unclonable function~\cite{iot2021}. 
     \item \textbf{Untrusted QAPs: } The proposed CR-QKD scheme relies on the trust of the intermediate QAPs. In this paper, we use techniques of trust computing to ensure that the the information stored in QAP is protected from external software attacks.  When a trusted platform is not available, designing a scheme that relaxes this assumption could also be a very good future research direction. 
 \item \textbf{Performance Optimization}: In this article, we presented a multi-user edge forwarding strategy, in which quantum keys were allocated as needed.  Unfortunately, its performance metrics, e.g., delay, secret key generation rate, and energy efficiency, are limited by those user pairs with weak channel reciprocity. How to optimize these performance metrics by allocating power or spectrum resources among different user pairs becomes an interesting topic and needs to be investigated. 
    \item \textbf {System Integration and Compatibility}: Our prototype was built on the USRP platform, which was different from commercial-off-the-shelf (COTS) devices. It is unknown whether these performances are still achievable on existing communication standards and whether CR-QKD will affect the network efficiency. 
More studies should be done on its system integration and compatibility issues, including frame format design, key management scheme and efficiency evaluation in practical communication systems.
\end{itemize}

\section{Acknowledgment}
We thank our colleagues Prof. Linning Peng, Mr. Yanjun Ding, Dr. Dong Wang, Mr. Siyun Wu and Dr. Xuyang Wang from the Purple Mountain Laboratories, for their help with the experimental platform. This work was supported in part by the National Key Research and Development Program of China under Grant 2020YFE0200600 and 2022YFB2902202, in part by the National Natural Science Foundation of China under Grant 62171121, in part by the Natural Science Foundation of Jiangsu Province under Grant BK20211160 and in part by Jiangsu Provincial Key
Laboratory of Network and Information Security under Grant BM2003201.
\bibliographystyle{IEEEtran}
\bibliography{IEEEabrv,ref}
\end{document}